\documentclass[conference]{IEEEtran}
\IEEEoverridecommandlockouts

\usepackage{cite}
\usepackage{amsmath,amssymb,amsfonts}
\usepackage{algorithmic}
\usepackage{graphicx}
\usepackage{textcomp}
\usepackage{xcolor}

\usepackage{makecell}
\usepackage{multirow}
\usepackage{adjustbox}
\usepackage{booktabs}
\usepackage{colortbl}
\definecolor{color1}{RGB}{0,180,0}
\definecolor{color2}{RGB}{180,0,0}
\definecolor{color3}{RGB}{0,0,180}
\definecolor{color4}{RGB}{250, 165, 0}
\definecolor{color5}{RGB}{128, 0, 128}
\usepackage{accessibility}
\def\BibTeX{{\rm B\kern-.05em{\sc i\kern-.025em b}\kern-.08em
    T\kern-.1667em\lower.7ex\hbox{E}\kern-.125emX}}

\usepackage{soul}
\usepackage{xcolor}

\usepackage{amsmath}
\begin{document}

\title{REVISION:Reflective Intent Mining and Online Reasoning Auxiliary for E-commerce Visual Search System Optimization\\
}
\edef\origskip{\the\skip\footins}
\author{\IEEEauthorblockN{Yiwen Tang\textsuperscript{*}, Qiuyu Zhao\textsuperscript{*}, Zenghui Sun, Jinsong Lan, Xiaoyong Zhu, Bo Zheng
\thanks{\textsuperscript{*} Equal contribution.}
}
\IEEEauthorblockA{\textit{Alibaba Group, China} \\
\{tangyiwen.tyw, xiaoyong.z, bozheng\}@alibaba-inc.com, \{zhaoqiuyu.zqy, zenghui.szh, jinsonglan.ljs\}@taobao.com}
}

\maketitle
\begin{abstract}

In Taobao e-commerce visual search, user behavior analysis reveals a large proportion of no-click requests, suggesting diverse and implicit user intents. These intents are expressed in various forms and are difficult to mine and discover, thereby leading to the limited adaptability and lag in platform strategies. 
This greatly restricts users’ ability to express diverse intents and hinders the scalability of the visual search system.
This mismatch between user implicit intent expression and system response defines the
\textit{User–SearchSys Intent Discrepancy}.
To alleviate the issue, we propose a novel framework \textbf{REVISION}.
This framework integrates offline reasoning mining with online decision-making and execution, enabling adaptive strategies to solve implicit user demands.
In the offline stage, we construct a periodic pipeline to mine discrepancies from historical no-click requests. 
Leveraging large models, we analyze implicit intent factors and infer optimal suggestions by jointly reasoning over query and product metadata. These inferred suggestions serve as actionable insights for refining platform strategies.
In the online stage, \textbf{REVISION-R1-3B}, trained on the curated offline data, performs holistic analysis over query images and associated historical products to generate optimization plans and adaptively schedule strategies across the search pipeline.
Our framework offers a streamlined paradigm for integrating large models with traditional search systems, enabling end-to-end intelligent optimization across information aggregation and user interaction. Experimental results demonstrate that our approach improves the efficiency of implicit intent mining from large-scale search logs and significantly reduces the no-click rate.

\end{abstract}


\section{Introduction}

Compared with text-based product search, e-commerce visual search is affected by a broader range of factors, including shooting angles, image quality, and user behavior patterns.
Taobao’s Pai Li Tao visual search system \cite{pailitao} alleviates these challenges through a modular pipeline integrating various specialized components for diverse user scenarios.
Despite the remarkable progress achieved under the paradigm, substantial no-click queries persist in large-scale systems, primarily due to the misinterpretation of user intent and the misalignment between retrieved products and implicit demands.

Empirical evidence indicates that user queries reflect diverse implicit intents beyond purchase-oriented goals.
However, conventional search systems exhibit limited intent coverage, constraining user-platform interaction to a single "visual matching" paradigm focused on image-to-image retrieval.
This results in suboptimal user experience and restricts the platform's functional diversification.
We identify the issue as \textit{User–SearchSys Intent Discrepancy}, attributed to inefficient mining of large-scale search logs and the predefined and constrained strategies in traditional visual search systems. 
We further outline two key challenges underlying the problem.




\textbf{Challenge 1. Mining and Discovery from Massive Data.} 
Existing methods for mining implicit intents from large-scale search logs rely on periodic manual annotation of sampled queries to identify potential mismatches. Annotators are guided by predefined tool lists from the visual search pipeline to select suitable tools for intent discrepancies. However, these approaches face critical scalability bottlenecks: manual annotation is inefficient, and the resulting limited data and handcrafted solutions cover only a narrow range of real user intents.
Attempts to replace manual annotation with algorithmic mining still depend on manually defined criteria, failing to fundamentally resolve the scalability issue.

\textbf{Challenge 2. Online Strategy Optimization.} 
Constrained by inefficient manual mining, existing visual search systems primarily leverage user-side information to enhance query features and employ rule-based strategies with discrete classifiers for fixed explicit intent optimization. 
These methods neglect implicit intent representation in user queries through historical search results, lacking global collaborative optimization. Moreover, they fail to support dynamic strategy adaptation required for handling multiple coexisting implicit intents within queries.

We thus ask: \textit{\textbf{Can we develop a framework that efficiently identifies discrepancies in historical requests and effectively addresses online user queries through Vision language model(VLM)\cite{vlm-survey}-based understanding and reasoning?}}
In response, we propose \textbf{REVISION}, first leveraging VLMs for two synergistic stages: efficient \textbf{offline} intent mining from historical requests and real-time \textbf{online} strategy optimization through flexible dispatch.

\textbf{1) Offline Stage.}
We design a periodically executed pipeline that aggregates similar no-click queries and associates them with actionable platform strategies. Based on Qwen2.5VL-72B~\cite{qwen25vl}, we extract visual features from queries and their corresponding retrieved products, integrate product attributes (price, title), and generate preliminary optimization suggestions for both user-side and platform-side strategies.
Furthermore, leveraging the advanced reasoning capability of 
Qwen3-30B-A3B~\cite{qwen3}, we refine these suggestions with detailed visual information, richer product attributes, and manually defined rules, yielding actionable optimization signals. The offline mining results provide insights for improving existing tools and inspire the design of optimization strategies.

\textbf{2) Online Stage.}
We convert offline optimization signals into trainable multiple-choice decision-making tasks. Inspired by Plan-Then-Execute\cite{Plan-Then-Execute}, REVISION-R1, built upon Qwen2.5VL-3B~\cite{qwen25vl}, is trained using offline mining data and suggestions to reason over real-time user query images and corresponding historical product results, dynamically predicting strategy optimization plans. The predicted actions select and execute appropriate downstream tools sequentially.

With this framework, we evolve the visual search system from simple image-to-image matching into an agentic architecture capable of collaborative multi-tool retrieval with implicit intent understanding.
We further design efficient online deployment strategies to seamlessly integrate our framework with the existing visual search systems.

The REVISION framework enables effective intent mining and discovery in the offline stage and performs global agentic optimization among the entire search pipeline in the online stage.
In the online A/B test, compared with previous pipeline, the ratio of no-click queries decreases by 13.91\% for trigger subset, while the Click-Through Rate (CTR), order volume, and Gross Merchandise Value (GMV) increase by 10.73\%, 13.60\%, and 10.73\%, respectively. Human evaluation further validates the effectiveness of our approach. These results demonstrate the practical potential and effectiveness of the proposed framework in e-commerce visual search systems.

\section{Related Works}
\textbf{Intent mining and understanding in Search and Recommender Systems.} 
Li J et al. employs Graph and Bert\cite{bert} for query classification.
Wang J et al.\cite{b2} integrates multiple dynamic behavior sequences and a dual-encoder to address the user intent understanding in cross-lingual search. J. Niu et al.\cite{b1} mines user preferences, user intentions, and potential relationships between items, and models them via contrastive learning to optimize user intent understanding in recommendation. L. Sang et al.\cite{b3} also performs intent optimization in Contrastive Learning (CL)-based recommender systems, leveraging a Heterogeneous Graph to incorporate intent information into the recommendation process. Sun K et al.\cite{b4} primarily address the problem of intent recognition in multi-modal scenarios by exploiting intra-video and cross-video context interactions to enhance intent modeling, while leveraging a cross-video bank for retrieval to mitigate error accumulation. New intent discovery leverages multi-task pre-training and contrastive learning with clustering~\cite{newintent}. S2TM~\cite{s2tm} integrate social and item semantics to handle sparsity and long-tailed distributions. AttnMix~\cite{attnmix} captures short-term intent via order-invariant session representations, while IOCLRec~\cite{IOCLRec} disentangles mixed intents through multi-granularity contrastive learning and dynamic segmentation. MIND~\cite{mind} infers intents using multimodal signals and LLMs by identifying shared attributes among co-purchased items.

\begin{figure*}[tb!]
  \centering
  \vspace{-8pt}
  \includegraphics[width=0.9\linewidth]{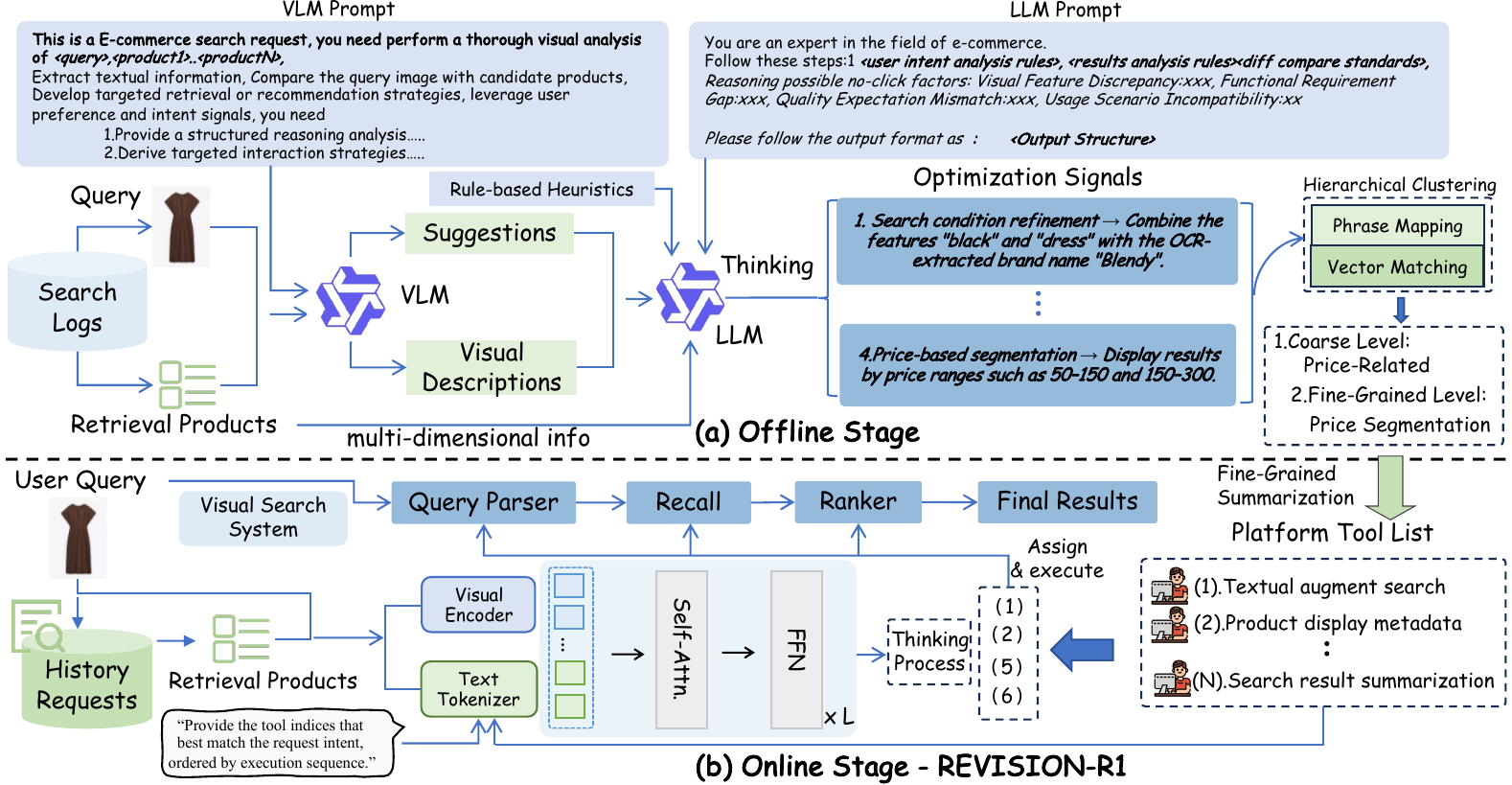}
  \vspace{-0.3cm}
  \caption{(a) illustrates the offline stage, where VLMs and LLMs are employed for reasoning, and hierarchical clustering is applied to organize the inferred results. The fine-grained clustering outcomes guide the expansion of the tool list. (b) shows the online stage, where the trained REVISION-R1-3B assigns executable tools to the visual search pipeline based on the query image, enabling targeted interventions across different stages of the process.}
\label{pipeline_1}
\vspace{-0.4cm} 
\end{figure*}

\textbf{VLM and LLM Applications in Search Systems.}
Several works~\cite{queryrewritinglargelanguage,query2doc} employ LLMs to refine user queries via rewriting, improving search effectiveness.
LAPS~\cite{laps} enhances conversational search personalization by modeling user preferences from past interactions. Pasa~\cite{pasa} utilizes dual LLM agents for query expansion and relevance evaluation in academic search.
BASES~\cite{bases} simulates user profiles and interactions using LLM-based agents tailored to specific personas, enabling realistic behavior modeling.
Baidu AI Search~\cite{baiduaisearch}, described by Yuchen Li et al., integrates multiple LLM agents in a collaborative framework to respond to user queries effectively.
OSrCIR~\cite{Reason-before-retrieve} uses reflective chain-of-thought to infer modification intents before retrieval, mitigating semantic drift. Chain-of-Intent~\cite{ChainI} combines Hidden Markov Models(HMMs) with LLMs to model intent transitions and synthesize dialogues.


Most related work differs fundamentally in task formulation, input/output specifications, and evaluation environment, focusing on behavior modeling or item ranking rather than system-level planning, targeting query rewrites instead of executable tool sequences, or using static datasets lacking multimodal result-aware feedback. 
Given these incompatible task paradigms, direct comparison remains impractical.
REVISION provides a unified reasoning framework that achieves greater data efficiency without large-scale pre-training or explicit user feedback, and effectively learns to infer what users need.
\section{Method}

\subsection{Motivation and Overview}

Figure~\ref{pipeline_1} illustrates the \textbf{REVISION} paradigm, comprising asynchronous offline and online stages.
In the offline stage, we leverage Qwen2.5VL and Qwen3 to jointly analyze historical query images and the corresponding retrieval products, extracting a series of actionable optimization signals.
Based on these signals, we perform hierarchical clustering via phrase mapping and vector similarity matching~\cite{sentence}. This yields multiple coarse-grained clusters, each further divided into fine-grained sub-clusters. These fine-grained groups indicate specific downstream tools, forming a curated tool list.
During online model training, we utilize the offline-generated data (a query paired with retrieval products, thinking process and ordered tool lists) as training samples. A VLM is trained to reason and predict appropriate tool sequences.
In the online stage, the trained model assigns executable tools to different search system stages based on real-time queries and historical retrieval products, optimizing outputs to mitigate ambiguity in request intent.
Required information for each tool is obtained from the query and products.

\subsection{Offline Stage}
The offline stage automatically mines users' latent intents from large-scale non-click search logs through multimodal understanding of historical queries and retrieved products, followed by structured reasoning over extracted visual signals.
\paragraph{\textbf{Analysis Steps of Large Models}}
The sampled data comprises a query image paired with multiple historical products, including corresponding titles, prices, and similarity scores to the query. We first input data into \textbf{Qwen2.5VL-72B} to extract visual information from the query and products. Without any manual constraints, the model then generates preliminary optimization suggestions, typically related to direct visual or semantic factors such as category, appearance, and price differences.
To enhance reasoning beyond visual understanding, we further employ \textbf{Qwen3-30B-A3B} as the text-based reasoning model. Beyond the outputs from Qwen2.5VL, we provide Qwen3 with abundant product metadata (e.g., origin, package size) and domain expert rules encoding visual search system biases. The former includes attributes such as origin and package size, and the latter introduces the inductive bias of the visual search system, encoded by domain experts in the form of rules. For instance, if the retrieval results show large price discrepancies concentrated in low ranges, it would be suggested to supplement with similar high-end products.
Integrating information from all four sources, Qwen3 generates a sequence of optimization signals in the format of \texttt{"Action -> Info"}, serving as valuable training data for downstream tools.
\texttt{Action} denotes the specific operation such as search condition refinement or price range segmentation; \texttt{Info} represents the information required to execute the action, e.g., visual features like ``black'' and OCR-extracted brand text ``Blendy'' for search conditions.

\paragraph{\textbf{Hierarchical Clustering Algorithm}}
Let $\mathcal{A}=\{a_i\}_{i=1}^{N}$ be short textual ``actions'' extracted from optimization signals. 
We build a two-level hierarchy over a predefined ontology of executable \emph{main categories} $\mathcal{C}$, each with \emph{subcategories} $\mathcal{S}_c$. 
Each $a_i$ is first preprocessed by removing punctuation and index markers, then tokenized with a word segmenter. 
For any label $\ell$ (either $c\in\mathcal{C}$ or $s\in\mathcal{S}_c$), we maintain a keyword synonym list $K_\ell$ used for robust lexical matching.
For action $a$ and label $\ell$, we compute a synonym-overlap score $s_{\text{syn}}(a,\ell)$ as the fraction of tokens in $a$ that match $K_\ell$ under substring equivalence, and a semantic score $s_{\text{sem}}(a,\ell)$ via cosine similarity between sentence-transformer embeddings of $a$ and the concatenation of $K_\ell$. 
We then form a combination
\[
s_\alpha(a,\ell)=\alpha\, s_{\text{syn}}(a,\ell) + (1-\alpha)\, s_{\text{sem}}(a,\ell),
\]
with $\alpha=0.7$ for main-category assignment and $\alpha=0.6$ for subcategory assignment.

\textit{\textbf{Level~1.}}
Each $a_i$ is assigned to 
\[
\arg\max_{c\in\mathcal{C}} s_{0.7}(a_i,c)
\]
if the maximal score exceeds a confidence threshold $\tau_1=0.40$; otherwise it is marked \emph{unassigned}. 
For unassigned items, we compute pairwise similarities and run DBSCAN\cite{dbscan} (Density-based clustering algorithm) on the precomputed distance matrix $d_{ij}=\max\{0,\,1-\cos(a_i,a_j)\}$ with $\varepsilon=0.5$ and $\text{min\_samples}=2$, yielding auxiliary semantic clusters.

\textit{\textbf{Level~2.}}
For main category $c$, the assigned actions are further partitioned over $\mathcal{S}_c$ using $s_{0.6}(a,s)$ and a relaxed threshold $\tau_2=0.35$; items below the threshold are routed to an ``other'' bucket under $c$.
Finally, the algorithm returns a hierarchical mapping (main category $\rightarrow$ subcategories $\rightarrow$ actions).
\paragraph{\textbf{Component-based Tools}}
Based on the executable signals from clustering, we extend and upgrade the existing tools and modularize them as standardized components for online scheduling integration(e.g. Display metadata adjustment component, Search result summarization component, External Textual search component and Target products labeling component). These components enable flexible composition via graphical configuration. The system dynamically constructs Directed Acyclic Graphs~\cite{DAG} for chain execution based on the planning list through automated configuration generation.

\subsection{Online Stage}
Trained on offline-mined data and suggestions, the online model captures real-time user intent and performs agentic tools invocation to optimize search results.
\paragraph{\textbf{Training Data Collection}}
We replace them with clustered sub-categories, transforming the original format from \texttt{Action~$\rightarrow$~Info} to \texttt{Sub-category~$\rightarrow$~Info}. The Sub-category pairs are extracted and mapped to ordered numeric labels, such as \texttt{(1)}, based on their positions in the tool list. 
As a result, the training data follows the structure:
\[
\texttt{Query + Retrieval Results} \rightarrow \texttt{(1) (3) (5)},
\]
which significantly improves controllability and reduces response time.
We observe that incorporating reasoning traces into the training data improves prediction accuracy. To achieve this, we leverage \textbf{Qwen3} to rephrase and compress the offline thinking process. The output format is structured as: \texttt{Thinking Process} + \texttt{(1) (3) (5)}.
At the tool level, the required execution information \texttt{Info} is extracted from the query and the retrieval products.

\paragraph{\textbf{REVISION-R1 Training}}
\textbf{\textit{Stage 1: Supervised Fine-Tuning.}}
We perform supervised fine-tuning using \textbf{Qwen2.5VL-3B}. The visual input $\mathcal{V}$ consists of a query image $I_q$ and up to $K=12$ retrieval product images $\{I_1, I_2, \dots, I_K\}$. The textual input $\mathcal{T}$ includes:
(1) Product metadata: structured strings, e.g., \texttt{Product $i$: price = 19, title = stylish black off-shoulder dress, quantity = 1}.
(2) Tool component details: represented by \texttt{Index}, \texttt{Title}, and \texttt{Description}. The description specifies the component functionality, along with its expected input and output formats.
The model is trained to reason and predict the tool components used for optimization, along with their execution order. The output sequence is denoted as $\mathcal{Y} = (y_1, y_2, \dots, y_T)$. We adopt the standard next-token prediction and optimize the cross-entropy loss:
\[
\mathcal{L}_{\text{CE}} = - \sum_{t=1}^{T} \log P(y_t \mid y_{<t}, \mathcal{V}, \mathcal{T}).
\]
\textbf{\textit{Stage 2: Reinforcement Learning.}}
To enhance the reasoning capability of the REVISION-R1, we build upon the existing GRPO~\cite{GRPO} algorithm and sample the hard training data used during the SFT stage~\cite{dapo} to guide the model to generate reasoning sequences autonomously. We formulate the model as a policy function $\pi_\theta$.
During training, for each query–retrieval list pair $(q, R)$, we sample $N$ candidate outputs $\{o_1, \dots, o_N\}$ from the current policy $\pi_\theta$. For each candidate $o_i$, we compute a task-specific reward $r_i = R(q, o_i)$, which is composed of multiple sub-reward components. We then calculate the group-normalized advantage:
\[
A_i = \frac{r_i - \text{mean}\{r_1, \dots, r_N\}}{\text{std}\{r_1, \dots, r_N\} + \delta},
\]
where $\delta$ is a small constant added for numerical stability.
Our reward function consists of two components:
\begin{itemize}
  \item \textbf{Format reward} ($r_\text{format}$): This encourages the model to produce structured reasoning sequences using specific format tokens such as \texttt{<think>...</think>}. A score of 1.0 is assigned if the \texttt{<think>} tag is used correctly; otherwise, the score is 0.
  
  \item \textbf{Answer accuracy reward} ($r_\text{ans}$): This measures the correctness of the final answer. A reward of 1.0 is assigned when either the predicted tool types or their execution order matches the ground truth, 2.0 when both match, and 0 otherwise.
\end{itemize}

\section{Experiments}
\begin{figure*}[tb!]
  \centering
  \vspace{-8pt}
  \includegraphics[width=0.94\linewidth]{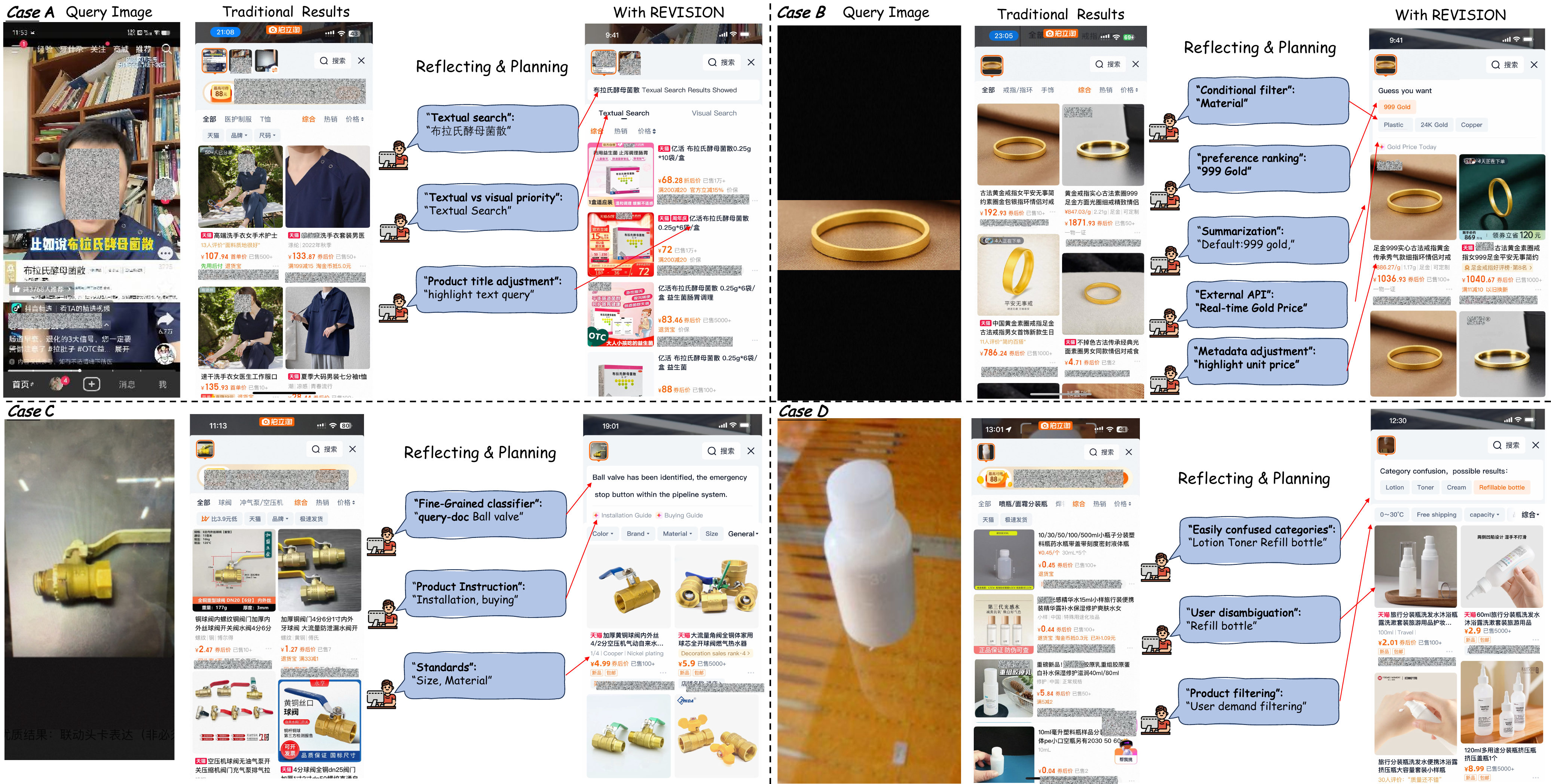}
  \vspace{-0.3cm}
  \caption{Online Comparisons of interface between traditional results and search system with REVISION. Some commercial and sensitive information of Query images and results is masked.Case A: REVISION identifies the user’s implicit need for medication information and prioritizes explanatory results over purely visual matches. Case B: REVISION restructures jewelry results using material- and price-aware signals to reduce cognitive load caused by visually similar but attribute-divergent products. Case C: REVISION highlights key specifications and functional details for standardized components, minimizing users’ verification effort. Case D: REVISION resolves ambiguity in visually unclear queries through preference-guided interaction.}
\label{case}
\vspace{-0.4cm} 
\end{figure*}
\subsection{Offline and Online Setups}
\paragraph{\textbf{Offline Mining pipeline details}}
Our offline mining stage is conducted on millions of unique image queries sampled weekly from historical no-click requests.
For Qwen2.5-VL, we limit the input product images to 12 per query.
Qwen2.5VL-72B is deployed on 4 PPU~\cite{alibaba_ppu_2025} (Alibaba-developed accelerator) GPUs.
For Qwen3-30B-A3B (deployed on 2 PPU GPUs), to reduce interference from irrelevant information, we rank the product metadata by importance and select the top 10 elements as input.
The VLM-LLM mining pipeline is powered by 60 PPU~\cite{alibaba_ppu_2025} GPUs.
This pipeline can process and analyze millions of query data within a single day.

In hierarchical clustering, the \emph{main categories} $\mathcal{C}$ and corresponding \emph{subcategories} $\mathcal{S}_c$ are automatically extracted by Qwen3 from the analysis step, not manually predefined. The clustering is scheduled weekly as part of the regular data processing pipeline.
We ensure that the \emph{main categories}, \emph{subcategories}, and keyword list $K_\ell$ are sufficiently large to provide broad coverage.
For large-scale and continuously updated data, incremental clustering efficiently incorporates newly generated data without reprocessing the entire dataset.

\paragraph{\textbf{Offline to Online Data Management}}
In the offline stage, We target no-click queries—image uploads without clicks within 30 seconds. After filtering bot traffic and low-quality images via CNN classifiers, we collect 8–12 million such queries daily from Taobao. The offline pipeline is orchestrated weekly via Airflow, with all intermediate artifacts stored in versioned partitioned Hive tables for traceability.

An Image Query Caching Module is deployed to reduce redundant processing of semantically similar image queries by reusing historical reasoning results. It combines a global image feature extractor with a vector database to detect near-duplicate queries (similarity \textgreater 0.85). Highly similar queries directly reuse cached global decisions while executing only downstream actions. Over time, this cache covers about 30\% of queries, achieving over 93\% accuracy and significantly reducing computation costs.
\paragraph{\textbf{Online deployment details}}
Our serving system is adapted from SGLang\cite{sglang}, deployed on a cluster of 45 PPU GPUs. We adopt the Prefill–Decode decoupled architecture\cite{pdsep} and design a real-time scheduling mechanism that monitors task load at the millisecond level, dynamically routing requests to low-load workers for maximum throughput. After generating the plan list, the system concurrently fetches required external data alongside the main search pipeline to minimize end-to-end latency. The additional latency is primarily due to the REVISION-R1 model in the triggered subset, contributing 95–100ms to {TP99~\cite{dean2013tail} (Latency at 99th percentile) and 45ms to average response time.
\paragraph{\textbf{Model Training Details}}
In the Supervised Fine-Tuning (SFT) stage, we collect 4.3M training samples by sampling hundreds of thousands from each main category, with thinking processes limited to 256 tokens.
For the Reinforcement Learning (RL) stage, we identify 680K difficult samples as training data where SFT predictions mismatch ground truth on the same 4.3M dataset.

\subsection{Evaluation and Ablation Study} 
\setcounter{table}{0}
\begin{table*}[tb!]
\centering
\begin{minipage}[t]{0.26\linewidth}
\centering
\vspace{-6pt}
\caption{Offline Stage Evaluation}
\vspace{-6pt}
\label{tab:evaluate}
\resizebox{\linewidth}{!}{%
\begin{tabular}{ccc|c}
\toprule
\textbf{\makecell{Metric}} & \textbf{Base Result} & \textbf{Test Result} & \textbf{\makecell{Diff}} \\
\midrule
\textbf{Top1 Relevance} & 28.57\% & 66.56\% & +37.99\% \\
\textbf{Top4 Relevance} & 36.38\% & 70.59\% & +34.21\% \\
\bottomrule
\end{tabular}%
}
\end{minipage}%
\hfill
\begin{minipage}[t]{0.37\linewidth}
\centering
\vspace{-6pt}
\caption{Online performance of A/B tests}
\vspace{-6pt}
\label{tab:abtest}
\resizebox{\linewidth}{!}{%
\begin{tabular}{cccccc}
\toprule
\textbf{\makecell{Online Metric}} & \textbf{CTR} & \textbf{CVR} & \textbf{\makecell{Order  Count}} & \textbf{GMV} & \textbf{\makecell{No-click Ratio}} \\
\midrule
\textbf{Trigger Subset} & +10.73\% & +8.82\% & +13.60\% & +10.73\% & -13.91\% \\
\textbf{General Set} & +0.44\% & +0.23\% & +0.58\% & +1.04\% & -0.92\% \\
\bottomrule
\end{tabular}%
}
\end{minipage}%
\hfill
\begin{minipage}[t]{0.32\linewidth}
\centering
\vspace{-6pt}
\caption{Ablation study on Mining Frequency}
\vspace{-6pt}
\label{tab:freq}
\renewcommand{\arraystretch}{0.85}
\resizebox{\linewidth}{!}{%
\begin{tabular}{lcccccc}
\toprule
 & \textbf{Hit Rate} & \textbf{CTR} & \textbf{CVR} & \textbf{GMV} & \textbf{\makecell{Non-click  Ratio}} & \textbf{\makecell{+GPUs}}\\
\midrule
T+4  & \textbf{31.1\%} & +0.41\% & +0.19\% & +0.97\% & -0.86\% & 35 \\
\rowcolor{gray!15} T+8 (Ours) & 30.4\% & \textbf{+0.44\%} & \textbf{+0.23\%} & \textbf{+1.04\%} & \textbf{-0.92\%} & \textbf{0} \\
T+16 & 29.8\% & +0.40\% & +0.22\% & +1.01\% & -0.90\% & 0 \\
\bottomrule
\end{tabular}%
}
\end{minipage}
\vspace{-15pt}
\end{table*}

\begin{table}[tb!]
\centering
\vspace{-5pt}
\caption{Benchmark results on intent reflection, with thinking content accuracy (Qwen3, CIDEr, BLEU-4, METEOR, ROUGE) and answer accuracy (Tool Match and Order Match)}
\vspace{-6pt}
\label{tab:benchmark}
\setlength{\tabcolsep}{2pt}
\scalebox{0.83}{
\begin{tabular}{lccccc|cc}
\toprule
\textbf{Model} & \textbf{Qwen3} & \textbf{CIDEr} & \textbf{BLEU-4} & \textbf{METEOR} & \textbf{ROUGE} & \textbf{\makecell{Tool \\ Match}} & \textbf{\makecell{Order \\ Match}} \\
\midrule
GPT-4o~\cite{gpt4o}        & 45.1  & 56.8  & 26.5 & 16.7 & 32.8 & 45.3 & 24.7 \\
Seed-1.5VL~\cite{seedvl}   & 36.0  & 49.1  & 13.8 & 18.2 & 24.7 & 42.6 & 27.8 \\
Gemini 2.5 Pro~\cite{gemini}  & 40.9 & 61.5  & 21.9 & 11.4 & 26.9 & 50.1 & 30.3 \\
\midrule
OmniSearch~\cite{omnisearch} & 54.6 & 70.2  & 36.0 & 19.6 & 51.9 & 58.8 & 39.4 \\
\textbf{REVISION-R1} & \textbf{67.0} & \textbf{91.7}  & \textbf{53.1} & \textbf{30.3} & \textbf{61.9} & \textbf{75.2} & \textbf{58.1} \\
-SFT & 45.3 & 70.1 & 37.9 & 26.0 & 42.5 & 55.9 & 37.6 \\
-RL & 53.9 & 83.2 & 46.0 & 28.5 & 59.6 & 61.0 & 43.4 \\
\bottomrule
\end{tabular}
}
\vspace{-10pt}
\end{table}

\begin{table}[tb!]
\centering
\caption{Hyperparameter Sensitivity Analysis in Offline Mining}
\vspace{-6pt}
\label{tab:hyperparameter_sensitivity1}
\scalebox{0.85}{
\begin{tabular}{l|l|cc|cc}
\hline
\multirow{2}{*}{\textbf{ }} 
& \multirow{2}{*}{\textbf{Configuration}} 
& \multicolumn{2}{c|}{\textbf{Hyperparameters}} 
& \multicolumn{2}{c}{\textbf{Relevance (\%)}} \\ \cline{3-6}
& & $\alpha_1 / \alpha_2$ & $\tau_1 / \tau_2$ & Top-1 & Top-4 \\ \hline
— & \textbf{Ours} 
& \textbf{0.7 / 0.6} 
& \textbf{0.40 / 0.35} 
& \textbf{66.56} 
& \textbf{70.59} \\ \hline
\multirow{2}{*}{\makecell[l]{\textit{Confidence} \\ \textit{Thresholds ($\tau$)}}} 
& High Precision 
& 0.7 / 0.6 
& 0.55 / 0.50 
& 63.12 
& 67.85 \\
& High Recall 
& 0.7 / 0.6 
& 0.30 / 0.25 
& 58.45 
& 62.90 \\ \hline
\multirow{2}{*}{\makecell[l]{\textit{Weighting} \\ \textit{Factors ($\alpha$)}}} 
& Syntax-Dominant 
& 0.9 / 0.8 
& 0.40 / 0.35 
& 60.23 
& 64.15 \\
& Semantic-Dominant 
& 0.4 / 0.3 
& 0.40 / 0.35 
& 61.88 
& 65.40 \\ \hline
\end{tabular}

}
\vspace{-10pt} 
\end{table}

\begin{table}[tb!]
\centering
\caption{Analysis of Online Reinforcement Learning Training.}
\vspace{-6pt}
\label{tab:online_sensitivity1}
\resizebox{\columnwidth}{!}{%
\begin{tabular}{l|cc|cc}
\hline
\multirow{2}{*}{\textbf{Configuration}} 
& \multicolumn{2}{c|}{\textbf{Hyperparameters}} 
& \multicolumn{2}{c}{\textbf{Performance (\%)}} \\ \cline{2-5}
& Group Size ($G$) & Reward Setup 
& Thinking Accuracy & Tool Match \\ \hline
\rowcolor{gray!10}
\textbf{Ours} 
& \textbf{8} 
& \textbf{$r_{format} + r_{ans}$} 
& \textbf{67.0} 
& \textbf{75.2} \\ \hline
\multicolumn{5}{l}{\textit{Group Size}} \\
Small Group 
& 4 
& $r_{format} + r_{ans}$ 
& 62.4 
& 70.8 \\
Large Group 
& 12 
& $r_{format} + r_{ans}$ 
& 67.3 
& 75.0 \\ \hline
\multicolumn{5}{l}{\textit{Reward Design}} \\
w/o Format Reward 
& 8 
& $r_{ans}$ only 
& 43.8 
& 48.2 \\
Binary Answer Reward 
& 8 
& $r_{format} + r_{binary}$ 
& 61.1 
& 68.4 \\ \hline
\end{tabular}
}

\vspace{-10pt} 
\end{table}

\begin{table}[tb!]
\centering
\caption{Ablation Study on Clustering Distance Metrics}
\vspace{-6pt}
\label{tab:metric_ablation1}
\footnotesize
\setlength{\tabcolsep}{4pt}
\scalebox{0.85}{%
\begin{tabular}{lccc}
\toprule
\textbf{TopN} & \textbf{Ours (Cosine)} & \textbf{Variant A ($L_2$)} & \textbf{Variant B ($L_1$)} \\
\midrule
Top-1 & \textbf{66.56} & 62.84 & 59.15 \\
Top-4 & \textbf{70.59} & 67.20 & 64.05 \\
\bottomrule
\end{tabular}
}
\vspace{-15pt}
\end{table}

\paragraph{\textbf{Offline Pipeline Evaluation}}
We randomly sampled 10,000 online queries that triggered optimization strategies and recruited 10 assessors with search ranking expertise. 
For each query, assessors reviewed the query image with professional annotations of ground-truth intent descriptions, then examined the top-1 and top-4 retrieved products from both the baseline system (traditional PaiLiTao pipeline) and the test system (our REVISION offline mining pipeline), including product images, titles, and prices. 
Effectiveness is measured by Top-1/Top-4 Relevance: assessors simulated user behavior to judge whether at least one highly relevant product aligned with the ground-truth intent appeared in the final ranked results.
As shown in Table~\ref{tab:evaluate}, REVISION's offline mining pipeline significantly outperformed the baseline, improving search quality by \textbf{37.99\%} in top-1 results and \textbf{34.21\%} in top-4 results. 
The inter-assessor agreement reached \textbf{91\%}, indicating high consistency.

\paragraph{\textbf{REVISION-R1 Evaluation}}
We designate the aforementioned annotated data as the test set, with ground truth from the offline stage comprising a reasoning process and tool indices in the form of labels such as (1), (3).
To measure the effectiveness of online intent capture, model performance is evaluated along two key dimensions:
First, \textbf{thinking content accuracy} (reasoning quality), assessed via LLM-as-a-judge (Qwen3) and natural language generation metrics (CIDEr\cite{cider}, BLEU-4\cite{belu}, METEOR\cite{meteor}, and ROUGE\cite{rouge}, standard text similarity metrics) to assess the validity of the reasoning content and its alignment with human-corrected  ground truth. 
Second, \textbf{answer accuracy} (tool-calling correctness), which is further divided into two aspects:
(1) exact/partial tool selection match, and (2) tool sequence order match.}
For baseline comparison, we select GPT-4o~\cite{gpt4o}, Seed-1.5VL~\cite{seedvl}, and Gemini 2.5 Pro\cite{gemini}, using the same set of prompts.
As shown in Table~\ref{tab:benchmark}, our REVISION-R1 significantly outperforms other models and demonstrates that both the SFT and RL training stages are indispensable.
In the thinking content evaluation, REVISION-R1 outperforms OmniSearch~\cite{omnisearch} by \textbf{13.6\%} on the Qwen3 metric.
In the answer accuracy evaluation, REVISION-R1 achieves \textbf{16.4\%} and \textbf{18.7\%} higher tool matching and order matching rates, respectively, compared with OmniSearch, which is a GPT-4V–based adaptive retrieval planning agent.

\paragraph{\textbf{Online A/B Test}}
To evaluate the real-world performancek, REVISION was deployed in Taobao Visual Search for an online A/B test.
We allocated 10\% of user traffic to each strategy to rigorously assess its effectiveness and stability. Optimized results displayed only for queries with low relevance scores, with limited daily cue frequency per user to ensure smooth experience.
As shown in Table~\ref{tab:abtest}, the test was conducted over seven days. The triggered subset achieved notable gains over the baseline, while the overall experiment group showed consistent positive improvements. Approximately 17\% of total traffic fell within the triggered subset, and all results were statistically significant (p\textless0.05).

\paragraph{\textbf{Ablation Study of Offline Mining Frequency}}
As shown in Table~\ref{tab:freq}, higher offline mining frequency (e.g., T+4) improves cache hit rate and query freshness by enriching intent cache with recent trends. However, it captures volatile, noise-driven signals rather than stable user behavior patterns, introducing unstable new intents that interfere with product recommendation and negatively impact user clicks and purchases, while incurring additional GPU costs. In mature search systems, user behavior evolves slowly, and valuable new intents require sufficient signal accumulation, clustering, manual validation, and A/B testing before deployment. Moderately reducing mining frequency (e.g., T+8) captures stable intent trends that better align with genuine user intents, generating more stable ranking results that effectively encourage user engagement and conversion, while using only online elastic resources without extra GPU cost, thereby achieving a better benefit–cost trade-off. Further reduction (e.g., T+16) yields no significant gains over T+8.

\paragraph{\textbf{Ablation Study of Data Engineering Aspects}}
We conduct ablation studies on critical data engineering components: (1) offline mining hyperparameters, (2) online training configurations, and (3) DBSCAN similarity metrics.

\textit{Offline Mining Hyperparameters.} Table~\ref{tab:hyperparameter_sensitivity1} shows our configuration ($\alpha=0.7/0.6$, $\tau=0.40/0.35$) achieves optimal balance. High $\tau$ reduces recall by filtering signals; low $\tau$ introduces noise. Balanced $\alpha$ combines lexical precision and semantic generalization, syntax-dominant settings miss LLM expression diversity, while semantic-dominant settings suffer drift. We use 12 input images balancing API cost and quality.

\textit{Online Training Hyperparameters.} Table~\ref{tab:online_sensitivity1} shows $N=8$ provides optimal performance-efficiency trade-off for GRPO group size. Smaller $N$ increases variance (-4.6\% thinking accuracy at $N=4$); larger $N$ yields marginal gains (+0.3\% at $N=12$) with high memory cost. Format reward $r_{format}$ is essential, removal disrupts reasoning chains (-27.0\% tool match). Graded answer rewards (0/1/2) outperform binary rewards by encouraging full correctness. Learning rate $1\times10^{-6}$ balances performance and stability.

\textit{DBSCAN Similarity Metrics.} Table~\ref{tab:metric_ablation1} shows Cosine distance significantly outperforms Euclidean ($L_2$) and Manhattan ($L_1$). Cosine captures angular similarity in Sentence-BERT embeddings, while magnitude-sensitive metrics fail to cluster semantically similar but magnitude-variant signals, reducing relevance (e.g., -3.72\% top-1 for Euclidean).

\section{Conclusion}
The paper introduces the implicit User–SearchSys Intent Discrepancy problem and proposes REVISION, a VLM-based agentic search framework integrating offline intent mining with online reasoning.
Offline, REVISION identifies and clusters intent discrepancies from large-scale search logs for strategy design. Online, it autonomously orchestrates tools for agentic optimization. A/B testing validates its effectiveness in alleviating intent discrepancies and enhancing user experience.
Beyond e-commerce, REVISION offers transferable insights for integrating large language models (LLMs) into search systems by leveraging data-driven workflows over handcrafted rules. It demonstrates that no-click interactions yield valuable signals when interpreted by reasoning models, with implications extending to recommendation and conversational systems.
Future work will unify offline traces with online fine-grained signals as memory and perception, propelling REVISION into a self-evolving agentic search system.



\end{document}